\documentclass[twocolumn,aps,showpacs,nofootinbib,prd]{revtex4-1}
\usepackage{amssymb}
\usepackage{amsmath}
\usepackage{graphics}
\usepackage{epsfig}
\usepackage[dvips]{color}

\newcommand\be{\begin{equation}}
\newcommand\ee{\end{equation}}
\newcommand\bes{\begin{eqnarray}}
\newcommand\ees{\end{eqnarray}}
\newcommand\ben{\begin{eqnarray*}}
\newcommand\een{\end{eqnarray*}}

\begin{document}

\title{Casimir Energy of an irregular membrane}

\author{B. Droguett}
\email{bdroguett@ucn.cl}
\affiliation{Departamento de F\'{\i}sica, Universidad Cat\'olica del Norte, Angamos 0610, Antofagasta, Chile}

\author{J.C. Rojas}
\email{jurojas@ucn.cl}
\affiliation{Departamento de F\'{\i}sica, Universidad Cat\'olica del Norte, Angamos 0610, Antofagasta, Chile}

\begin{abstract}
We compute the Casimir energy which arises in a bi-dimensional surface due to the quantum fluctuations of  a scalar field. We assume that the boundaries are irregular and the field obeys Dirichlet  condition. We re-parametrize the problem to one which has flat boundary conditions and the irregularity is treated as a perturbation in the Laplace-Beltrami operator which appears. Later, to compute the Casimir energy, we use zeta function regularization. It is compared the results coming from perturbation theory  with the WKB method.
\end{abstract}
\maketitle


\section{Introduction}

The Casimir effect can be considered among the few macroscopic manifestation of quantum phenomena. As it is well known, the Casimir effect originally appeared as a relative force between conductor or semiconductor surfaces due to the quantum fluctuations of vacuum \cite{casimir}, \cite{polder}. The Casimir effect is an interdisciplinary subject, which plays an important role in Quantum Field Theory, Condensed Matter, Gravitation and Mathematical Physics \cite{elizalde}, \cite{odintsov} . In  particular, for the subject of membrane theory, one can refer to \cite{od1}, \cite{od2}. 

In our work, we consider a problem in 2+1 dimensions, with a bi-dimensional surface bounded by an irregular border, where the Casimir force arise due to the quantum fluctuations of a free scalar field \cite{bordag}. We consider this problem as the case of an idealized membrane where the phonon fluctuations are responsible of the Casimir force between borders.

We work in Euclidean space and the spatial coordinates are re-parametrized, in order to convert the irregular borders in two parallel plates, so, the scalar dynamic is given by the resulting Laplace-Beltami operator, due to the coordinate transformation.

In order to compute the Casimir energy at zero temperature and its resulting force, we use the zeta function regularization as it is done in \cite{kirsten}. Were it is necessary to compute the determinant of the Laplace-Beltrami operator which arises from the integration of the scalar fields. For the sake of simplicity, we consider a rectangular shape of size $L \times a$ with Dirichlet boundary condition. So, zero modes are avoided. Of course, we consider the “length” $L$ much bigger than the “width” $a$, as it is usually done. 

The free energy in terms of zeta function, is given by

\[
E_{cas}=\frac{1}{2}\left[FP\zeta(-1/2) +\rm Res\zeta(-1/2)\ln(\mu)\right], 
\]

\noindent as it is shown in \cite{kirsten}. It means that the residues of the zeta function carry an ambiguity in the determination of the free energy, since it appears an arbitrary scale $\mu$, which is harmless if the residue does not depend on the parameter $a$, the separation. But, if it appears a dependence on $a$, it means that the method is not enough to determine the force on the borders. So it is necessary to try another approach.

We first use perturbation theory, and later, the WKB method \cite{bender},\cite{mckane}. Implying that we can compare both methods.  We conclude that the dominant terms are not the same when $L \rightarrow \infty$. But, if we consider $L$ finite, there appear differences in the contribution for the energy. We assume that the better method is the WKB, since it is also useful for obtaining the residues of the zeta function, which are related to the geometry of the system \cite{jeffres}.


\section{General setting}\label{zeta}

\subsection{Laplacian of a non regular object}
We consider a nearly rectangle membrane, 

A free scalar field on the membrane obeys the Laplace equation

\be
\label{L-B}
-\bigtriangleup_{s}\phi=\lambda\phi.
\ee

\noindent If we assume an irregular boundary, we can rescale the irregular lenght in order 
to obtain a rectangular boundary, so the dynamic of the scalar field is given by the Laplace-Beltrami 
operator

\be
\label{1}
\bigtriangleup_{s}=\frac{1}{\sqrt{g}}\frac{\partial}{\partial x^{i}}\left(\sqrt{g}g^{ik}\frac{\partial}{\partial x^{k}}\right).
\ee

\noindent If we consider two dimensions; x and y, where $0 \leq x \leq L $  and  $0\leq y \leq H(x)$. We assume  $a \ll L$, with
\begin{equation}
\label{H}
H(x)=a+h(x) \;\; {\rm and} \;\; h(x) \ll a.
\end{equation}\\

\noindent Rescaling x and y, 

\be
\label{Newvariables}
x=uL ,\qquad y=H(u )v,
\ee

\noindent where $u$, $v$ are fixed coordinates,  

\begin{equation}
\label{dominio}
0\leq u\leq 1,   \qquad   0\leq v\leq 1.
\end{equation}

\noindent We end up with a nondiagonal spacial metric tensor

\begin{equation}
\label{metrica}
g_{ik}=\begin{bmatrix}
L^{2}+(v h^{'}(u))^{2} & (a+ h(u))v h^{'}(u)  \\
 (a+ h(u))v h^{'}(u) & (a+ h(u))^{2}\\
\end{bmatrix},
\end{equation}\\

\noindent whose determinant

\begin{equation}
\label{jaja}
g=|\det(g_{ik})|=(a+ h(u))^{2}L^{2},
\end{equation}

\noindent help us to compute the contravariant metric tensor

\begin{equation}
\label{metgik}
g^{ik}=\begin{bmatrix}
\frac{1}{L^{2}} & -\frac{v h'(u)}{(a+ h(u))L^{2}} \\
 -\frac{v h'(u)}{(a+ h(u))L^{2}}  & \frac{L^{2}+(v h'(u))^{2}}{L^{2}(a+ h(u))^{2}}\\
\end{bmatrix}.
\end{equation}

\noindent With the above parametrization, the Laplace Beltrami operator over the fields is given by

\begin{widetext}

 \begin{eqnarray}-\bigtriangleup_{s}\phi&=&-\frac{1}{L^{2}}\frac{\partial^{2}\phi}{\partial u^{2}}-\left[\frac{L^{2}+(v h^{'}(u))^{2}}{(a+ h(u))^{2}L^{2}}\right]\frac{\partial^{2}\phi}{\partial v^{2}}-\frac{1}{L^{2}}\left[-\frac{v h^{''}(u)}{a+ h(u)}+\frac{2v( h^{'}(u))^{2}}{(a+ h(u))^{2}}\right]\frac{\partial \phi}{\partial v}\nonumber\\&&
+\frac{2 v h^{'}(u)}{(a+ h(u))L^{2}}\frac{\partial^{2}\phi}{\partial u\partial v}=\lambda\phi.\label{Ecu.autovalores}
\end{eqnarray}\\

\end{widetext}

\noindent So, to compute the determinant, we have to solve the eigenvalue equation

\[
-\bigtriangleup_{s}\phi= \lambda \phi.
\]

\noindent We shall assume $1\gg h(u)$, $h(u) \gg h'(u)/L$ and $h(u)\gg h''(u)/L$, so, keeping to the quadratic terms in $h(u)$, we end up with

\bes
\label{nablac}
-\bigtriangleup_{s}\phi &=& -\frac{1}{L^{2}}\frac{\partial^{2}\phi}{\partial u^{2}}
-\frac{1}{a^2} \frac{\partial^2 \phi}{\partial v^2}
\nonumber \\
&&- \left(
\frac{2 h(u)}{a^3}
-\frac{3 h(u)^2}{a^4}
\right)
\frac{\partial^{2}\phi}{\partial v^{2}}.
\ees

\noindent In order to perform perturbation theory, we identify

\bes
\label{potencial}
V(u,v)\phi &=&
 \left(
-\frac{2 h(u)}{a^3}
+\frac{3 h(u)^2}{a^4}
\right)
\frac{\partial^{2}\phi}{\partial v^{2}}
\nonumber \\
&\equiv & G(u) \frac{\partial^{2}\phi}{\partial v^{2}}.
\ees

\noindent For the sake of simplicity, we shall use Dirichlet boundary condition in both
coordinates:

\bes
 \phi_{n,m}(0,v) &=& \phi_{n,m}(1,v)=0,
\nonumber \\
  \phi_{n,m}(u,0) &=& \phi_{n,m}(u,1)=0.
\nonumber
\ees

\bigskip


\section{Perturbation theory}\label{perturbation}

First, we solve the standard Laplace equation

\begin{equation}
\label{ecuacionhomogenea}
-\bigtriangleup_{s}\phi^{0}=-\frac{1}{L^{2}}\frac{\partial^{2}\phi^{0}}{\partial u^{2}}-\frac{1}{a^{2}}\frac{\partial^{2}\phi^{0}}{\partial v^{2}}=\lambda^{0}\phi^{0},
\end{equation}

\noindent whose solution and eigenvalues are given by

\begin{eqnarray}
\label{funcion0}
\phi^{0}_{n,m}(u,v) &=& 2 \sin(m\pi u) \sin(n\pi v),
\\
\label{lambda0}
\lambda^{0}_{n,m} &=& \left(\frac{n\pi}{a}\right)^{2}+\left(\frac{m\pi}{L}\right)^{2},
\end{eqnarray}

\noindent since we are using Dirichlet boundary conditions, $n$ and $m$ are
positive integers

\[
\label{cond1}
1\leq n < \infty, \;\;  1\leq m<\infty.
\]

\noindent So, the first order correction term is 

\begin{eqnarray}
\label{coretion}
\delta \lambda^{p}_{n,m} &=& \int_{0}^{1}\int_{0}^{1} \phi^{0,*}_{n,m} V(u,v)\phi^{0}_{n,m}\,dudv
\nonumber \\
&=& \int_{0}^{1}\int_{0}^{1} \phi^{0,*}_{n,m} G(u) \frac{\partial^{2}\phi_{n,m}}{\partial v^{2}}dudv,
\end{eqnarray}

\noindent giving the eigenvalue

\begin{equation}
\label{lambdacos}
\lambda_{n,m}=\left(\frac{n\pi}{a}\right)^{2}+\left(\frac{m\pi}{L}\right)^{2}+
\pi^{2}n^{2} \int^{1}_{0}  G(u) \,du,
\end{equation}\\

\noindent where $h(u)$, introduced in (\ref{H}), shall be considered as a periodic function in the variable $u$.

The generalized zeta function is given by

\begin{equation}
\label{zettotal}
\zeta(s)=\dfrac{1}{\Gamma(s)}\int^{\infty}_{0} t^{s-1}\sum^{\infty}_{m=1}\sum^{\infty}_{n=1}e^{-\alpha m^{2}t-\beta n^{2}t}\,dt,
\end{equation}

\noindent where the parameters $\alpha$ and $\beta$ are

\begin{eqnarray}
\label{alphaa}
\alpha &=& \left(\frac{\pi}{L}\right)^{2}, \nonumber \\
\beta &=& \left(\frac{\pi}{a}\right)^{2}+\pi^2
\int^{1}_{0} G(u)\,du.
\end{eqnarray}

\noindent In order to obtain the thermodynamical potential, it is necessary to 
compute the generalized zeta function for $s=-1/2$

\bes
\label{evaluatezeta}
\zeta(-1/2)& =& \frac{1}{24}(\sqrt{\beta}+\sqrt{\alpha})-\frac{\zeta_{R}(3)}{8\sqrt{\alpha\beta}\pi^{2}}(\beta^{3/2}+\alpha^{3/2})
\nonumber \\
&-&\frac{\pi}{4\sqrt{\alpha\beta}}\sum^{\infty}_{n=1}\sum^{\infty}_{m=1}\left(\frac{\pi^{2} n^{2}}{\beta }+\frac{\pi^{2} m^{2}}{\alpha }\right)^{-3/2},
\ees

\noindent we end with an expression for the Casimir energy

\begin{widetext}

\be
E_{cas}  \simeq  \frac{1}{48}(\sqrt{\beta}+\sqrt{\alpha})-\frac{\zeta_{R}(3)}{16\sqrt{\alpha\beta}\pi^{2}}(\beta^{3/2}+\alpha^{3/2})
+\frac{\pi^{-1}}{8\sqrt{\alpha\beta}}{\beta}^{3/2} \left( -\alpha\sqrt {\beta}\sqrt {{\frac {\beta
+\alpha\,{\pi }^{2}}{\alpha}}}+\alpha\,{\pi }\sqrt {\beta}-\,
\sqrt {\alpha}\beta \right).
\ee

\end{widetext}

\noindent For the leading term in the Limit $\omega=0$ and $L \rightarrow \infty$, we have the energy and force per unit lenght given by

\begin{equation}
\label{energia0}
\frac{E_{cas}}{L} \equiv \frac{E_0}{L}=-\frac{\zeta_{R}(3)}{16\pi a^{2}},
\end{equation}

\begin{equation}
\frac{ F_{0}}{L}=-\frac{\zeta_{R}(3)}{8\pi a^{3}}.
 \end{equation}

\noindent If we assume  a regular behavior for $h$, in order to parametrize it as a trigonometric 
function

\be
\label{formacos} 
h(u)=\epsilon \cos(\omega u+\phi), \;\; \omega= \delta L,
\ee

\noindent assuming $\delta\ll 1$,  $\epsilon \ll 1$ and $ \epsilon \ll \delta$.
In such case, the energy is given by

\begin{widetext}

\bes
  E_{cas} & =&-{\frac {a \zeta_{R}(3) }{16\pi \,{L}^{2}}}
+{\frac {{\pi }^{4}}{8{a}^{3}L}}
+{\frac {\pi }{48 L}}
-{\frac {{\pi }^{3}\sqrt {{L}^{2}+{\pi }^{2}{a}^{2}}}{8 {a}^{4}L}}
-{\frac {\zeta_{R}(3)L }{16\pi {a}^{2}}}
+{\frac {\pi }{48 a}}
-{\frac {{\pi }^{3}}{8 {a}^{4}}}
\nonumber \\
&+&\left(
-{\frac {\zeta_{R}(3) \sin( \omega) \cos( \phi) }{16 \omega\pi {L}^{2}}}
-{\frac {3 {\pi }^{4}\sin( \omega) \cos( \phi) }{8\omega {a}^{4}L}}+\cdots
\right)\epsilon
+\left(\frac{\pi}{64a^{3}}-\frac{3\zeta_{R}(3)L}{32\pi a^{4}}+\cdots
\right)\epsilon^{2}.
\ees

\end{widetext}

\bigskip

\noindent In the limit $L \rightarrow \infty$, we have 

\be
E_{cas}=-{\frac {\zeta_{R}(3)L }{16\pi {a}^{2}}}
-\frac{3\zeta_{R}(3)L}{32\pi a^{4}} \epsilon^2.
\ee

\noindent For the case where we have two irregular surfaces parametrized by

\begin{equation}
 \label{forma2}
 h(u)=\epsilon_{2}\cos(\omega u+\phi)-\epsilon_{1}\cos(\omega u), 
 \end{equation}

\noindent with  $\omega=\delta L$,  $\delta \ll 1$ and
$\epsilon_{1},\epsilon_{2} \ll 1$. The energy we obtain, is the following

\begin{equation}
\label{cas2}
 E_{cas}=-\frac{\zeta_{R}(3)L}{16\pi a^{2}}-\frac{3\zeta_{R}(3)}{32\pi a^{4}}(\epsilon^{2}_{1}+\epsilon^{2}_{2}-2\epsilon_{1}\epsilon_{2}\cos(\phi))L.
 \end{equation}\\

\bigskip


\section{WKB method}\label{wkb}

\bigskip

We can rewrite (\ref{nablac})

\begin{equation}
\label{ñeñe}
-\bigtriangleup_{s}\phi=-\frac{1}{L^{2}}\frac{\partial^{2}\phi}{\partial u^{2}}-	Q(u)\frac{\partial^{2}\phi}{\partial v^{2}}=\lambda\phi.
\end{equation}\\

\noindent Where, $Q(u)$ is given by

 \begin{equation}
 \label{forma}
 Q(u)=\frac{1}{a^2}-G(u)=\frac{1}{a^2}- \left(
-\frac{2 h(u)}{a^3}
+\frac{3 h(u)^2}{a^4}
\right),
  \end{equation}\\

\noindent  with $G(u)$ defined in (\ref{potencial}). If we use the following Ansatz

\[
\phi(u,v)=M_{n, \lambda}(u)\sin(n\pi v),\;\;n \in \mathbb{N^{+}},
\]

\noindent it leads us to the equation 

\bigskip

\begin{equation}
\label{M}
M_{n,\lambda}^{''}(u)+(\lambda-n^{2}\pi^{2}Q(u))L^{2}M_{n,\lambda}(u)=0.
\end{equation}

\noindent The eigenvalues of the sistem are given by the zeros of $M_{n,\lambda}(1)$.  We define the normalized function as it is done in \cite{jeffres}

\[
D_n(\lambda)=\frac{M_{n,\lambda}(1)}{M_{n,0}(1)},
\]

\noindent so, the zeta function can be expressed as

\begin{equation}
\label{zetak}
\zeta(s)=\sum_{n=1}^{\infty}\frac{1}{2\pi i}\int_{\Gamma}\lambda^{-s}\frac{d\ln D_{n}(\lambda)}{d\lambda}\,d\lambda.
\end{equation}\\

\noindent Choosing an appropiate contour integral, where $\lambda \rightarrow iz$, we have

\begin{equation}
\label{zetazeta}
\zeta(s)=\frac{\sin(\pi s)}{\pi}\sum_{n=1}^{\infty}(\pi n)^{-2s}\int_{0}^{\infty} z^{-s}
\frac{d\ln D_{n}(z)}{d z}\,d z.
\end{equation}

Since the contour has been rotated, we replace $\lambda=-n^{2}z\pi^{2}$ in (\ref{M}),

\begin{equation}
\label{M2}
M_{n,z}^{''}(u)-n^{2}L^{2}\pi^{2}(z+Q(u))M_{n, z}(u)=0.
\end{equation}

\noindent We can look for a solution of the form

\begin{equation}
\label{WKB}
M_{n, z}(u)=e^{\int_{0}^{u} S(\sigma,z,n) \;d\sigma },
\end{equation}
 
\noindent where $S(u,z,n)$ obeys

\begin{equation}
 \label{EcS}
 S^{2}(u,z,n)+\frac{\partial S(u,z,n)}{\partial u}=n^{2}L^{2}\pi^{2}\left(z+Q(u)\right).
 \end{equation}

\noindent We search a solution  in powers of $n^{-1}$

\begin{equation}
 \label{S}
 S(u,z,n)=\sum_{i=-1}^{N}a_{i}(u,z)n^{-i}.
 \end{equation}

\noindent So, we can obtain the coefficients $a_i$ recursively. The first three terms are

\bes
a_{-1}(u,z) &=& \pm\pi L\sqrt{z+Q(u)},  \nonumber \\
a_{0}(u,z) &=& -\frac{1}{2a_{-1}(u,z)}\frac{\partial a_{-1}(u,z)}{\partial u},
\nonumber \\
a_{1}(u,z) &=& -\frac{1}{2 a_{-1}(u,z)}
\left(\frac{\partial a_{0}(u,z)}{\partial u}+a_{0}^{2}(u,z)\right) \!\!,
\ees

\noindent and so on.

\bigskip

The general solution is given by

\begin{equation}
  \label{MAB}
  M_{n,z}(u)= A e^{\int_{0}^{u} S_+(\sigma,z,n) \;d\sigma }
+B e^{\int_{0}^{u} S_-(\sigma,z,n) \;d\sigma },
  \end{equation}

\noindent where $S_{\pm}$ is the splitting 

\begin{equation}
  \label{Smm}
  S_{\pm}(u,z,k)=\pm S_{1}+S_{2},
  \end{equation}

\noindent with

\bes
 \label{zvm}
 S_{1}(u,z,n) &=& a_{-1}(u,z)n+\frac{a_{1}(u,z)}{n}+\frac{a_{3}(u,z)}{n^{3}}+\cdots,
\nonumber  \\
 S_{2}(u,z,n) &=&  a_{0}(u,z)+\frac{a_{2}(u,z)}{n^{2}}+\frac{a_{4}(u,z)}{n^{4}}+\cdots .
 \ees

\noindent Since we are dealing with Dirichlet boundary conditions, we have

\begin{equation}
\label{condM}
M_{n,z}(0)=M_{n,z}(1)=0 \;\; {\rm and} \;\; M_{n,z}^{'}(0)=1.
\end{equation}

\noindent We end with the solution for $M_n$

\bes
\label{Mnorm}
M_{n,z}(u) &=& \frac{ e^{\int_{0}^{u}S_{1}(\sigma,z,n)\,d\sigma}}{2\sqrt{S_{1}(1,z,n)S_{1}(0,z,n)}} 
\nonumber \\ && \times
\left(1-e^{-2\int_{0}^{u}S_{1}(\sigma,z,n)\,d\sigma}\right).
\ees

\noindent The term in (\ref{zetazeta}) reads

 \begin{equation}
  \label{drog}
  \ln D_{n}(-n^{2}z)=\ln\left(\frac{M_{n,z}(1)}{M_{n,0}(1)}\right),
  \end{equation}

\noindent so

\bes
\ln D_{n}(-n^{2}z) &=& \int_{0}^{1} S_{1}(\sigma,z,n)\,d\sigma
\nonumber \\
&& -\frac{\ln \left(S_{1}(0,z,n)\right)+\ln \left(S_{1}(1,z,n)\right)}{2}
 \nonumber \\
&&
 +\ln \left(1- e^{-2\int_{0}^{1}S_{1}(\sigma,z,n)\,d\sigma}\right)
\nonumber \\ &&
-\ln\left(2 M_{n,0}(1) \right).
 \ees

\noindent Expanding $S_1$

 \bes
\ln(S_{1}(u,z,n)) &=& \ln(a_{-1}n)+\frac{a_{1}}{a_{-1}n^{2}}
+\frac{a_{3}}{a_{-1}n^{4}}
\nonumber \\ 
&-& \frac{1}{2}\left(\frac{a_{1}}{a_{-1}n^{2}}+\frac{a_{3}}{a_{-1}n^{4}}\right)^{2}+\cdots,
\ees

\noindent the derivative of the previous expression assumes the form

\begin{equation}
 \label{dlnz}
 \frac{\partial\ln S_{1}(u,z,n)}{\partial z}=\sum_{i=0}^{\infty}b_{2i}(u,z)n^{-2i}.
\end{equation}  

\noindent The generalized zeta function can be expressed as

\begin{widetext}
\bes
\label{zetag}
\zeta(s) &=& \frac{\sin(\pi s)}{\pi}\int^{\infty}_{0}z^{-s}\int^{1}_{0}\sum^{\infty}_{i=0}\pi^{-2s}\zeta_{R}(2s+2i-1)\frac{\partial a_{2i-1}(\sigma,z)}{\partial z}\,d\sigma\,dz
\nonumber \\
&&-\frac{\sin(\pi s)}{2\pi}\int^{\infty}_{0}z^{-s}\sum_{i=0}^{\infty}(b_{2i}(1,z)+b_{2i}(0,z))\pi^{-2s}\zeta_{R}(2s+2i)\,dz
\nonumber \\
&&+ \frac{\sin(\pi s)}{\pi}\sum_{n=1}^{\infty}(n\pi)^{-2s}\int_{0}^{\infty}\,dz z^{-s}\frac{2e^{-2\int_{0}^{1}S_{1}(\sigma,z,n)\,d\sigma}}{1-e^{-2\int_{0}^{1}S_{1}(\sigma,z,n)\,d\sigma}}\frac{\partial}{\partial z}\int_{0}^{1}S_{1}(\sigma,z,n)\,d\sigma.
 \ees

\end{widetext}

\noindent Since the free energy is given by 

\[
E_{cas}=\frac{1}{2}\left[FP\zeta(-1/2) +\rm Res\zeta(-1/2)\ln(\mu)\right],  
\]

\bigskip

If we choose a regular boundary of the form

\[ 
h(u)=\epsilon\cos(\omega u+\phi),
\]

 \noindent  with $ \omega=\delta L$, $\delta\ll  1$ and 
$ \epsilon \ll 1$. From (\ref{zetag}), we obtain

\begin{widetext}

 \bes
 \rm FP\zeta(-1/2) &=& -{\frac {\zeta_R (3) L}{8\,\pi {a}^{2} }}
+{\frac {
\pi }{24\, a}}+\frac{\pi}{24\,L}
+\left(\frac{\zeta_{R}(3)\sin(\omega)\cos(\phi)L}{4\,\pi \omega a^{3}}-\frac{\pi\cos(\phi)}{48\,a^{2}}
 +\frac{\omega\sin(\omega)\cos(\phi)}{a\pi L}+...\right)\epsilon
\nonumber \\
&&+\left(-\frac{3\zeta_{R}(3)L}{16\,\pi a^{4}}+ 
\frac{3\omega\cos(\phi)\sin(\phi)}{16\,\pi a^{2}L}+\cdots \right)\epsilon^{2},
\ees

\end{widetext}

 \noindent and the residue

\bes
\rm Res\zeta(-1/2) &=& \left({\frac {{\omega}^{2}
\cos( \omega+\phi) }{64 \, \pi L^{2} }} +\cdots
\right)\epsilon
\nonumber\\
&+& \frac {15 \omega^2}{1024 \,\pi a L^2}
\left( \sin(2 \omega)\sin(2\phi) \right.
\nonumber \\
&&
\left. -2\cos^2(\omega)\cos^2(\phi)
+\cdots
\right) \epsilon^{2},
\ees

\bigskip

\noindent implying a Casimir energy of the form (leading with $L \rightarrow \infty$) 

\begin{equation}
\label{energiasegunda}
E_{cas}=E_0
-\frac{3\zeta_{R}(3)L}{32\,\pi a^{4}}\epsilon^{2}
+\frac{\delta^{2}L}{32\,\pi a^{2}} \epsilon^{2},
\end{equation}\\

\noindent with $E_0$, defined in (\ref{energia0})

\[
\label{energia00}
E_{0}=-\frac{\zeta_{R}(3)L}{16\pi a^{2}}.
\]

\noindent 
The Casimir energy when the two surfaces are irregular is given by, 

\bes
E_{cas} &=& -\frac{\zeta_{R}(3)L}{16\,\pi a^{2}}-\frac{3\zeta_{R}(3)}{32\,\pi a^{4}}(\epsilon^{2}_{1}+\epsilon^{2}_{2}-2\epsilon_{1}\epsilon_{2}\cos(\phi))L
\nonumber \\
&& +\frac{\delta^{2}}{32\, \pi a^{2}}(\epsilon^{2}_{1}+\epsilon^{2}_{2}-2
\epsilon_{1} \epsilon_{2} \cos(\phi))L+\cdots ,
 \ees

\noindent it does not coincide with the term obtained in (\ref{cas2}).


\section{Conclusions}\label{conclusions}

In this work, we computed the free energy of a bi-dimensional spatial surface with irregular boundaries, where the physics is played by a scalar field. We assumed a rectangular shape with the width $a$ much smaller than the length $L$ of the surface. For the sake of simplicity, we assume Dirichlet boundary condition on the borders. First, we used perturbation theory and after, the WKB method. 

In general, we find that perturbation theory does not coincide with the WKB method, for finite $L$. In the limiting case
$L\rightarrow \infty$, there is also a discrepancy with perturbation theory.
We conclude that perturbation theory, at least at first order, is not enough to describe the physics of the system.


\bigskip

{\bf Acknowledgments}\\[.2cm]
B.D. and J.C.R. aknowledge the support of  FONDECYT under grant No. 1120770 and J.C.R. aknowledges support of FONDECYT under grant No. 1095217. We thank Alejandro Ayala, Marcelo Loewe and Cristi\'an Villavicencio for useful commets and suggestions.

\end{document}